\documentclass[letterpaper, 10pt, conference]{ieeeconf}

\let\labelindent\relax

\usepackage{amsmath,amssymb,paralist,subfigure,cite,graphicx,color,amsbsy,float}

\usepackage{stmaryrd,mathrsfs,url}
\usepackage{cuted,mathtools,lipsum}
\usepackage[ruled,vlined,linesnumbered]{algorithm2e}

\usepackage{pifont}
\usepackage[utf8]{inputenc} 
\usepackage[T1]{fontenc}    
\usepackage{hyperref}       
\hypersetup{
    colorlinks=true,
    linkcolor=cyan,
    filecolor=mnodea,      
    urlcolor=cyan,
    citecolor=lime,
}
\usepackage{url}            
\usepackage{booktabs}       
\usepackage{amsthm}     
\usepackage{nicefrac}       
\usepackage{lipsum}
\usepackage[sc]{mathpazo}   
\usepackage{dsfont}
\usepackage[dvipsnames]{xcolor}

\usepackage{baskervillef}
\usepackage{pgfplots}
\pgfplotsset{compat=newest}
\usetikzlibrary{plotmarks}

\newtheorem{lem}{Lemma}

\newtheorem{rem}{Remark}

\newtheorem{cor}{Corollary}

\def\ve{\varepsilon}

\def\mb{\mathbf}

\def\mc{\mathcal}

\IEEEoverridecommandlockouts                
\begin{document}

\title{\LARGE \bf Distributed CPU Scheduling Subject to Nonlinear Constraints}

\author{Mohammadreza Doostmohammadian, Alireza Aghasi,  Apostolos I. Rikos, Andreas Grammenos, \\   Evangelia Kalyvianaki, Christoforos N. Hadjicostis, Karl H. Johansson,  Themistoklis Charalambous
\thanks{M. Doostmohammadian and T. Charalambous are with the School of Electrical Engineering at Aalto University, Finland, Email: \texttt{firstname.lastname@aalto.fi}. A. Aghasi is with Georgia State University, GA, USA, email: \texttt{aaghasi@gsu.edu}. 
A. Grammenos is with the Department of Computer Science and Technology, University of Cambridge, Cambridge, and the Alan Turing Institute, 	London, UK. E-mail: \texttt{ag926@cl.cam.ac.uk}. E. Kalyvianaki is with the Department of Computer Science
and Technology, University of Cambridge, Cambridge, UK. Email: \texttt{ek264@cl.cam.ac.uk}. C. N. Hadjicostis and T. Charalambous are with the Department of Electrical and Computer Engineering, University of Cyprus, Cyprus, email: \texttt{\{chadjic,charalambous.themistoklis\}@ucy.ac.cy}. A. I. Rikos and K. H. Johansson are with the Division of Decision and Control Systems, KTH Royal Institute of Technology, Sweden, email: \texttt{\{rikos,kallej\}@kth.se}. }
\thanks{ This work is supported in part by
the European Commission through the H2020 Project Finest Twins under Agreement 856602.}}

\maketitle
\thispagestyle{empty}

\begin{abstract}
	This paper considers a network of collaborating agents for local resource allocation subject to nonlinear model constraints.  In many applications, it is required (or desirable) that the solution be anytime feasible in terms of satisfying the sum-preserving global constraint. Motivated by this, sufficient conditions on the nonlinear mapping for anytime feasibility (or non-asymptotic feasibility) are addressed in this paper. For the two proposed distributed solutions, one converges over directed weight-balanced networks and the other one over undirected networks. In particular, we elaborate on uniform quantization and discuss the notion of $\ve$-accurate solution, which gives an estimate of how close we can get to the exact optimizer subject to different quantization levels. This work, further, handles general (possibly non-quadratic) strictly convex objective functions with application to CPU allocation among a cloud of data centers via distributed solutions. The results can be used as a coordination mechanism to optimally balance the tasks and CPU resources among a group of networked servers while addressing quantization or limited server capacity.
	
\end{abstract}

\section{Introduction}
Allocation of resources and utilities over a multi-agent network is  considered in this paper. This problem finds application in different control scenarios ranging from coverage control and task allocation  to electricity power scheduling \cite{cherukuri2015tcns,cherukuri2016initialization,8062536,vrakopoulou2017chance}. The general idea is to optimally determine the allocated amount of resources from a fixed total among a group of users or agents. Recently, the emergence of Internet-of-Things (IoT) has motivated distributed solutions over networks, where agents locally solve the problem in their neighborhood with no direct knowledge of distant agents or global information. In many large-scale applications, localized processing, and cloud computing motivate such \textit{distributed} resource allocation strategies instead of traditional centralized solutions. Example applications include managing the balance between energy resources and energy demand over the smart grid, allocating the fixed amount of tasks over a multi-agent network, or assigning the amount of computing load to the network of data servers \cite{makridis2020robust,rikos2021optimal,kalyvianaki2009self}. In the context of resource
management in Cloud infrastructures, we particularly focus on the latter application where some networked data centers (computing nodes) need to be assigned by CPU cycles (resources) in a distributed fashion. The total sum of resources is limited and fixed and the computing nodes follow a distributed algorithm to locally balance the CPU utilization by local information-exchange with other nodes. In general, in CPU scheduling the jobs are allocated in quantized (or discrete) values. Further, other than quantized CPU allocation and in general applications, the data-sharing setup is typically involved with bandwidth efficiency and limited capacity concerns, and thus, mandates quantized information exchange. This quantization issue needs to be addressed in general networked scenarios.

\subsection{The problem}  
The problem of sum-preserving resource allocation is in the following standard form,
\begin{eqnarray} \label{eq_dra}
\min_\mb{x}
~ & F(\mb{x}) = \sum_{i=1}^{n} f_i(\mb{x}_i)\\ \nonumber
& \text{s.t.} ~~ \sum_{i=1}^{n} \mb{x}_i = b,~~
& \mb{x}_i \in \mc{X}_i
\end{eqnarray}
with $\mb{x}_i,b \in \mathbb{R}$, $f_i: \mathbb{R} \rightarrow \mathbb{R}$, and $\mc{X}_i \subseteq \mathbb{R}$ representing a range of admissible values for states $\mb{x}_i$. The latter represents the so-called \textit{box constraints} for $\mb{x}_i \in \mathbb{R}$ in the form $\mb{x}_i \in [m_i~ M_i]$.
As discussed later, the problem can be extended to the case where $\mb{x}_i \in \mathbb{R}^{d_i}$ and $\mc{X}_i \subseteq \mathbb{R}^{d_i}$ where the \textit{local constraints} are defined in the form \cite{mikael2021cdc},  
\begin{equation} \label{eq_local_const}
\mc{X}_i = \{\mb{x} \in \mathbb{R}^{d_i}: g_i^j(\mb{x}) \leq 0, j=1,\dots,p_i \}
\end{equation}
with $g_i^j:\mathbb{R}^{d_i} \rightarrow \mathbb{R}$ as convex and twice-differentiable functions on $\mc{X}_i$. In general, the sum-preserving global constraint can be also of higher-order form with $\mb{x}_i,b \in \mathbb{R}^m$.

Among the existing solutions, other than the classic linear ones \cite{cherukuri2015tcns,cherukuri2016initialization,8062536,boyd2006optimal}, the work by \cite{mikael2021cdc} suggests a \textit{local reallocation} optimization algorithm at every iteration to address all-time feasibility. On the other hand, there exist many primal-dual solutions that do not guarantee primal-feasibility (or anytime-feasibility), but instead asymptotically reach feasibility \cite{aybat2016distributed,nedic2018improved}. Many existing works focus on linear solutions with ideal communication and actuation at the node dynamics. However, in reality, multi-agent systems (e.g., mobile robotic networks, connected generators over the smart grid, or collaborating distributed data centers) are subject to practical nonlinearities. For example, the shared information for task/CPU scheduling among data centers (or servers) are quantized \cite{rikos2021optimal} or  robot actuators performing coverage allocation are subject to saturation \cite{doostmohammadian2009novel}. The work by \cite{magnusson2018communication} further addresses the notion of $\ve$-accuracy over star multi-agent networks, i.e., the number of communication bits needed to reach the $\ve$-neighborhood of the exact optimizer. In the same line of research, Ref. \cite{taes2020finite} considers unconstrained distributed optimization via single-bit information-exchange over limited-capacity communication networks. Other than the mentioned nonlinearities imposed by the nature of the actuation and communication, other kinds of nonlinearities are added for the purpose of improving the convergence rate or to reach the optimal value in (prescribed) fixed-time \cite{garg_cdc20} or finite-time \cite{taes2020finite}. These further motivate the nonlinear model consideration in this paper.   

\subsection{Main Contributions}  
In this paper, the main contributions are: (i) we address possible nonlinearities in the dynamics of the agents due to imperfect actuation and limited communication capabilities. This is motivated, for example, by limited and/or quantized range of action in actuators and, similarly, possible clipping and quantization in communication channels. Other node-based and link-based nonlinearities are further applicable to address, for example, robustness to disturbances and pre-defined (or fixed) convergence time. Some examples regarding nonlinear \textit{consensus} protocols are discussed in \cite{mrd20211st,mrd2020fast,wei2018nonlinear}. In this paper, we discuss convergence subject to both sector-based and non-sector-based nonlinearities, for example, logarithmic quantization and uniform quantization. (ii) We show exact convergence under sector-based nonlinearities, while for uniform quantization (as an example of non-sector-based nonlinearity) we prove convergence to the $\ve$-neighborhood of the optimizer. In the latter case, the concept of $\ve$-accuracy is considered. This notion implies the quantization level to ensure reaching $\ve$-neighborhood of the optimal point. On the other hand, for a given quantization level (or the number of bits) one can address the best $\ve$-accurate solution that can be achieved while satisfying the feasibility constraint at all times. 
In particular, (iii) we discuss the application in resource allocation 
and CPU scheduling over networked servers \cite{rikos2021optimal}. (iv) Unlike some works restricted to quadratic costs \cite{rikos2021optimal}, this work can address general strictly (and strongly) convex cost functions (possibly non-quadratic) due to, e.g., the use of different barrier functions and penalty functions addressing the local constraints to advance the quadratic cost model in \cite{rikos2021optimal}. The results can further address different types of practical nonlinearities imposed on the coordination mechanism among the servers, for example, saturated capacity, quantization scheme of different sizes, and fast sign-based solutions.  Further, (v) we advance the assumption in \cite{rikos2021optimal,mikael2021cdc} by considering uniform-connectivity over time instead of all-time connected networks.

\subsection{Some Preliminary Concepts} \label{sec_pri}
Following the Karush-Kuhn-Tucker (KKT) condition, the following lemma finds the condition on the optimizer $\mb{x}^*$ as the solution of \eqref{eq_dra}. Define the gradient vector $\nabla F = [\partial_{x_1} f_1(\mb{x}_1);\dots;\partial_{x_n} f_n(\mb{x}_n)]$. 

\begin{lem} The optimizer $\mb{x}^*$ as the solution of \eqref{eq_dra} is in the form $\nabla F \in \mbox{span}(\mb{1}_n)$, i.e., $\partial_{x_j} f_j(\mb{x}_j^*) = \partial_{x_i} f_i(\mb{x}_i^*)$ for all $i,j$. 
\end{lem}
See the proof and more details in \cite{mrd20211st,mrd2020fast}. Note that the above lemma holds for the equality-constraint problem \eqref{eq_dra} without local constraints \eqref{eq_local_const}. The box constraints ($d_i=1$) are addressed by additive penalty terms discussed later in Section~\ref{sec_penalty}. 
One can reformulate the problem and extend it to  \textit{weighted}-sum-preserving constraints as follows, 
\begin{eqnarray} \label{eq_dra2}
\min_\mb{y}
~ & \widetilde{F}(\mb{y}) = \sum_{i=1}^{n} \widetilde{f}_i(\mb{y}_i)\\ \nonumber
& \text{s.t.} ~~ \sum_{i=1}^{n} a_i\mb{y}_i = b,~~
& \mb{y}_i \in \mc{Y}_i
\end{eqnarray}
By change of variable in the form $a_i \mb{y}_i=\mb{x}_i$, the above problem takes the form \eqref{eq_dra} and follows similar solution. Notice that $a_i$s need to satisfy composition conditions \cite[Section~3.2.4]{boyd2004convex} to ensure convexity of the local sets $\mc{X}_i$s after change of variables (as a composition of $\mc{Y}_i$s and linear transformation $a_i \mb{y}_i=\mb{x}_i$). 

\subsection{The Assumptions} \label{sec_ass}
The following assumptions on the cost functions hold throughout the paper:
\begin{enumerate}
	\item The local cost functions $f_i$ are strictly (or strongly) convex and smooth\footnote{For the proof of convergence only strict convexity is used. In order to determine the \textit{rate of convergence} $v$-strongly convex assumption is adopted.}. 
	\item The feasible solution set of problem~\eqref{eq_dra} is non-empty and compact.
\end{enumerate}
The first assumption allows to address the unique optimizer via KKT conditions and is widely considered in the literature. The second assumption is particularly challenging if there are different local constraints  $\mb{x}_i \in \mc{X}_i$ and the combination of these $\mc{X}_i$s and the sum-preserving constraint $\sum_{i=1}^{n} \mb{x}_i = b$ needs to be feasible. Algorithms are proposed in \cite{cherukuri2015tcns,mikael2021cdc} to render feasible initialization for such cases. 

The following assumptions (for the proof of convergence) hold on possible nonlinearities on the agents' dynamics: 
\begin{enumerate} [(i)]
	\item The nonlinearities satisfy  $0<\underline{\alpha} \leq \frac{h(z)}{z} \leq \overline{\alpha}$ (sector-based), i.e., they are strongly sign-preserving and monotonically non-decreasing nonlinear mapping.
	\item $h(z)$ is an odd mapping, i.e., $h(-z)=-h(z)$ and $h(0)=0$. 
\end{enumerate}

The following are standard assumptions on the multi-agent network (or the graph topology) in the consensus literature:
\begin{enumerate} [(I)]
	\item The network is undirected with symmetric weights. 
	\item The network is uniformly-connected or B-connected, i.e., the union of the networks over every time-interval $B$ is connected.  
\end{enumerate}

Note that for some special cases we relax the assumption~(I) to general \textit{weight-balanced directed networks}. In terms of network connectivity, Assumption~(II) advances existing solutions \cite{mikael2021cdc,magnusson2018communication} to dynamic (possible disconnected) networks, i.e., the cases for which the network might be disconnected during some time instances but their union is connected over a finite time interval $B$. This occurs in mobile multi-agent applications with limited communication resources where the links over the network come and go as the agents (e.g., robots) move in and out of the communication range of the other agents.    

\subsection{Paper Organization}
The rest of the paper is as follows. Section~\ref{sec_nonlin} introduces the distributed solutions subject to possible  nonlinearities. In Section~\ref{sec_quant},  the convergence of uniform quantization (as a non-sector-based nonlinearity) and the notion of $\ve$-accurate solution are discussed. Section~\ref{sec_sim} provides an example application in CPU scheduling and related simulations. Finally, Section~\ref{sec_con} concludes the paper.

\section{Nonlinear Distributed Solutions} \label{sec_nonlin}
Two nonlinear distributed gradient-Laplacian solutions are considered in this paper. The continuous-time (CT) solutions are in the form, 
\begin{align} \label{eq_ct_nonlin1}
\dot{\mb{x}}_i &=  \eta\sum_{j\in \mc{N}_i}W_{ji}(t)h(\partial_{x_j} f_j(t)-\partial_{x_i} f_i(t)), \\ \label{eq_ct_nonlin2}
\dot{\mb{x}}_i &=  \eta \sum_{j\in \mc{N}_i}W_{ji}(t)(h(\partial_{x_j} f_j(t))-h(\partial_{x_i} f_i(t))),
\end{align} 
The CT solutions find application, e.g., in economic dispatch problem and power generation scheduling, see \cite{mrd20211st,mrd2020fast}. In discrete-time (DT),

\small \begin{align} \label{eq_dt_nonlin1}
\mb{x}_i(k+1) &= \mb{x}_i(k) + \overline{\eta} \sum_{j\in \mc{N}_i}W_{ji}(k)h(\partial_{x_j} f_j(k)-\partial_{x_i} f_i(k)), \\ \label{eq_dt_nonlin2}
\mb{x}_i(k+1) &= \mb{x}_i(k) + \overline{\eta} \sum_{j\in \mc{N}_i}W_{ji}(k)(h(\partial_{x_j} f_j(k))-h(\partial_{x_i} f_i(k))),
\end{align} \normalsize
with $h(\cdot)$  representing possible node-based or actuation nonlinearity (protocols \eqref{eq_ct_nonlin1} and \eqref{eq_dt_nonlin1}) or link-based or communication nonlinearity (protocols \eqref{eq_ct_nonlin2} and \eqref{eq_dt_nonlin2}) at the agents' dynamics. This nonlinear function could be either (i) imposed by the nature of the agents' dynamics, e.g., due to control saturation and/or quantization, or (ii) added purposefully by the designer, e.g., to improve the convergence rate and/or robustness properties with respect to noise and disturbances by using sign-based solutions. 

\begin{lem} [Convergence] \label{lem_conv}
	Let the assumptions in Section~\ref{sec_ass} hold. The continuous-time solutions \eqref{eq_ct_nonlin1}-\eqref{eq_ct_nonlin2} and discrete-time solutions \eqref{eq_dt_nonlin1}-\eqref{eq_dt_nonlin2} converge to the exact optimizer $\mb{x}^*$ as the solution of problem \eqref{eq_dra}.
\end{lem}

The detailed proof for convergence and uniqueness of the solution under CT dynamics  \eqref{eq_ct_nonlin1}-\eqref{eq_ct_nonlin2} are given in \cite{mrd20211st,mrd2020fast} assuming general strictly convex cost functions. The proof can be extended to the DT case using the following lemma.

\begin{lem} \label{lem_strict}
	Let Assumptions~(1)-(2) hold. Consider two points $\mb{x}_1, \mb{x}_2 \in \mathbb{R}^n$, and $\delta \mb{x} := \mb{x}_1-\mb{x}_2$.  There exist $0<\alpha<1 $ and $\widehat{\mb{x}} = \alpha \mb{x}_1 + (1-\alpha)\mb{x}_2$  such that,
	\begin{align}
	F(\mb{x}_1) = F(\mb{x}_2) + \nabla F(\mb{x}_2)^\top \delta \mb{x}+  \frac{1}{2} \delta \mb{x}^\top \nabla^2 F(\widehat{\mb{x}}) \delta \mb{x}. \label{eq_taylor}
	\end{align}
	Then, 
	\begin{align} \label{eq_taylor_1}
	F(\mb{x}_1) \geq F(\mb{x}_2) + \nabla F(\mb{x}_2)^\top \delta \mb{x} +  v\delta \mb{x}^\top \delta \mb{x},
	\\
	F(\mb{x}_1) \leq F(\mb{x}_2) + \nabla F(\mb{x}_2)^\top \delta \mb{x} +  u\delta \mb{x}^\top \delta \mb{x}.
	\label{eq_taylor_2}
	\end{align}
\end{lem} 

Define the Lyapunov function as the residual $\overline{F}(k) = F(\mb{x}(k))-F(\mb{x}^*)$. For two consecutive (feasible) states $\mb{x}(k+1), \mb{x}(k)$ define ${\delta \mb{x}(k) := \mb{x}(k+1)-\mb{x}(k)}$.
To satisfy $\overline{F}(k+1) \leq \overline{F}(k)$, from Lemma~\ref{lem_strict} one can prove that, 
\begin{align} \label{eq_proof1}
\nabla F^\top \delta \mb{x}  + u \delta \mb{x}^\top \delta \mb{x}  \leq 0.
\end{align} 

Recall that for a weight-balanced connected graph $\mc{G}$ and its associated Laplacian matrix $L_g = D-W$ with $D = \mbox{diag}[\sum_{j\in \mc{N}_i}W_{ji}] = \mbox{diag}[\sum_{j\in \mc{N}_i}W_{ij}]$, define $\lambda_n,\lambda_2$ as the largest and smallest non-zero eigenvalue of $L_g$. For $\mb{x} \in \mathbb{R}^n$ and ${\overline{\mb{x}} := \mb{x} - \frac{\mb{1}_n^\top \mb{x}}{n} \mb{1}_n}$,
\begin{align} \label{eq_laplac}
\mb{x}^\top L_g \mb{x} &= \overline{\mb{x}}^\top L_g \overline{\mb{x}},
\\      \label{eq_laplace}
\lambda_2 \|\overline{\mb{x}} \|_2^2 \leq \mb{x}^\top & L_g \mb{x} \leq \lambda_n \|\overline{\mb{x}} \|_2^2
\end{align}

Using \eqref{eq_laplac}-\eqref{eq_laplace} and substituting $\delta \mb{x}$ from Eq. \eqref{eq_dt_nonlin1}-\eqref{eq_dt_nonlin2}, further assume strongly convex functions  
satisfying $2v\leq \partial_x^2 f_i(\mb{x}_i) \leq 2u$ and sector-based nonlinearities satisfying $\underline{\alpha} \leq \frac{h(\mb{z})}{\mb{z}} \leq \overline{\alpha}$. Then, similar Lyapunov analysis as in \cite{mrd20211st,mrd2020fast}, one can prove convergence for any step-rate $\overline{\eta}>0$ satisfying,
\begin{eqnarray} \label{eq_eta}
\overline{\eta} \leq \dfrac{2\underline{\alpha}\lambda_2} {u\lambda_n^2\overline{\alpha}}.
\end{eqnarray}  
Then, the linear convergence rate follows as,
\begin{eqnarray} \label{eq_cov_rate}
\frac{\overline{F}(k+1)}{\overline{F}(k)} \leq 1- \overline{\eta}v(\underline{\alpha}\lambda_2 - \frac{u}{2}\lambda_n^2\overline{\alpha} \overline{\eta}).
\end{eqnarray} 
The proof can be easily extended to B-connected graphs with $L_g$  as the Laplacian matrix of the union graph over the time-interval $B$, i.e., considering $\frac{\overline{F}(k+B)}{\overline{F}(k)}$ in the above formula. See \cite{mrd20211st,mrd2020fast} for more information.  

\begin{rem} 
	In problem~\eqref{eq_dra2}, following the KKT conditions, the optimizer satisfies $\nabla \widetilde{F}(\mb{y}^*) \in \mbox{span}(\mb{a})$. 
\end{rem}

\subsection{The Local Constraints} \label{sec_penalty}
The local constraints $\mb{x}_i \in \mc{X}_i$ can be addressed via adding penalty functions \cite{bertsekas_lecture} or barrier functions \cite{mikael2021cdc} to the local costs $f_i$. Some commonly used penalty functions to address the box-constraints are discussed here. The cost function is updated as, 
\begin{equation}
f_i^c (\mb{x}_i) = f_i(\mb{x}_i) + c [\mb{x}_i - M_i]^+ + c [m_i - \mb{x}_i]^+
\end{equation}
with $[u]^+ = \max \{u,0\}$ and $c>0$ penalizing the deviation from the admissible range of values. It is known that the solution of this penalized case can become arbitrary close to the exact optimizer by choosing $c$ sufficiently small \cite{nesterov2013introductory}. This non-smooth function can be substituted by the following smooth equivalents \cite{nesterov2013introductory,csl2021},
\begin{eqnarray} \label{eq_equiv1}
L(u,\mu)= &=& \frac{1}{\mu} \log (1+\exp (\mu u)) \\ \label{eq_equiv2}
[u]^{+\kappa} &=& ([u]^{+})^\kappa, ~\kappa>1,\kappa \in \mathbb{N} 
\end{eqnarray}
It can be shown that the maximum gap between the two functions  $[u]^+$ and \eqref{eq_equiv1} inversely scales with $\mu$, i.e.,
$$L(u,\mu)-[u]^{+} \leq \frac{1}{\mu}$$
and the two can become arbitrarily close by selecting $\mu$ sufficiently large~\cite{slp_book}. In general, for local constraints in the form \eqref{eq_local_const}, the penalty functions can be written as $c \sum_{j=1}^{p_i} [g_i^j(\mb{x})]^+$.
Similarly, some barrier functions $\mc{B}_i^j(\mb{x}_i)$ are proposed in the literature \cite{bertsekasnonlinear,mikael2021cdc} to be added to the local costs in the form $f_i^c (\mb{x}_i) = f_i(\mb{x}_i) + c \sum_{j=1}^{p_i} \mc{B}_i^j(\mb{x}_i)$. Following \eqref{eq_local_const}, $\mc{B}_i^j(\mb{x}_i)$ is defined real valued for $\mb{x}_i \in \mc{X}_i$, i.e., $g_i^j(\mb{x}_i)<0$, and following Assumption (1), 
\begin{enumerate}
	\item The barrier function needs to be convex and smooth.
	\item If $g_i^j(\mb{x}_i) \rightarrow 0^-$ (i.e., the function approaching zero from negative values), then $\mc{B}_i^j(\mb{x}_i) \rightarrow  \infty$.
\end{enumerate}
Some standard example barrier functions are given as \cite{bertsekasnonlinear},
\begin{eqnarray} \label{eq_bar1}
\mc{B}_i^j(\mb{x}_i) &=& - \log (-g_i^j(\mb{x}_i)) \\ \label{eq_bar2}
\mc{B}_i^j(\mb{x}_i) &=& \frac{-1}{g_i^j(\mb{x}_i)} 
\end{eqnarray}
These are respectively known as logarithmic and inverse barrier functions.

\subsection{The global Constraint: Anytime Feasibility}
As mentioned in the introduction, many applications mandate solution feasibility at all times, i.e., the global constraint $\sum_{i=1}^{n} \mb{x}_i = b$ hold at all times along the solution dynamics. This implies that at any termination time, the resulting outcome $\mb{x}$ of the proposed anytime-feasible protocols \eqref{eq_ct_nonlin1}-\eqref{eq_dt_nonlin2} satisfy $\sum_{i=1}^{n} \mb{x}_i = b$. In application, e.g., the economic dispatch problem, this means that the produced power and the demand are balanced at all times to avoid system break-down \cite{mikael2021cdc,cherukuri2015tcns}. Similarly, in balancing the CPU utilization among a group of data centers, the algorithm needs to be feasible at all times such that the allocated CPU resources meet the workloads required by the servers \cite{rikos2021optimal,kalyvianaki2009self}.   
 
\begin{lem} [Anytime Feasibility] \label{lem_feasible_intime}
	Suppose that Assumption~(2), Assumption~(ii), and Assumption~(I) hold. By any feasible initialization, the state of agents remain feasible under the CT dynamics~\eqref{eq_ct_nonlin1}-\eqref{eq_ct_nonlin2} for all $t>0$ and under the DT dynamics~\eqref{eq_dt_nonlin1}-\eqref{eq_dt_nonlin2} for all $k\geq 1$.
\end{lem}

The proof for CT case over uniformly-connected undirected graphs is discussed in \cite{mrd20211st,mrd2020fast}. For the DT case, the proof similarly follows. First, note that from Assumption~(2), the feasible solution exists. For protocol \eqref{eq_dt_nonlin1}, 
\begin{align} \nonumber
\sum_{i=1}^n \mb{x}_i&(k+1) = \sum_{i=1}^n \mb{x}_i(k) \\
&+ \overline{\eta} \sum_{i=1}^n \sum_{j\in \mc{N}_i}W_{ji}(k)h(\partial_{x_j} f_j(k)-\partial_{x_i} f_i(k)), 
\end{align}

Following Assumption~(ii) and Assumption~(I), the last term is equal to zero. This is because for two neighboring agents $i,j$, we have $W_{ij}=W_{ji}$ and 
$$h(\partial_{x_j} f_j(k)-\partial_{x_i} f_i(k)) = -h(\partial_{x_i} f_i(k)-\partial_{x_j} f_j(k)).$$

The feasibility proof of \eqref{eq_dt_nonlin2} for undirected graphs similarly follows. For link-based nonlinearities \eqref{eq_ct_nonlin2} and \eqref{eq_dt_nonlin2} one can extend the proof even to weight-balanced directed graphs. 

\begin{cor} \label{cor_wb}
	For protocols \eqref{eq_ct_nonlin2} and \eqref{eq_dt_nonlin2} over a weight-balanced graph, 	
\begin{align} \nonumber
	\sum_{i=1}^n &\mb{x}_i(k+1) = \sum_{i=1}^n \mb{x}_i(k) \\
	&+ \overline{\eta} \sum_{i=1}^n \sum_{j\in \mc{N}_i}W_{ji}(k)h(\partial_{x_j} f_j(k))-h(\partial_{x_i} f_i(k)), \label{eq_wb_prov} 
\end{align}
\end{cor}

Recall that for a weight-balanced graph $\mc{G}$ and its associated Laplacian matrix $L_g$,
we have
$\mb{1}_n^\top L_g \mb{z} = 0$,
where $\mb{z} \in \mathbb{R}^n$ and $\mb{1}_n$ as the vector of $1$s. Now considering $\mb{z} = [h(\partial_{x_1} f_1(k));\dots;h(\partial_{x_n} f_n(k))]$, the last term in \eqref{eq_wb_prov} is zero and Corollary~\ref{cor_wb} follows.  




\section{Quantization and $\ve$-Accuracy} \label{sec_quant}
In this section, we compare the convergence for two cases: sector-based nonlinearities satisfying Assumption~(i)-(ii), and sign-preserving 
(but not strongly) odd nonlinear mapping. Note that the main difference of the two cases is that for the second case $ \dfrac{dh}{dx}(0)=0$ while for the first case $ \dfrac{dh}{dx}(0)>0$. In particular, we consider logarithmic quantization versus uniform  quantization respectively as  examples of the first and second case. Following Lemma~\ref{lem_conv}, for sector-based nonlinearities the exact convergence is achieved, i.e., substituting the strongly sign-preserving function $h(z)=\mbox{sgn}(z)\exp\left(q\left[\dfrac{\log(|z|)}{q}\right] \right)$ 
in \eqref{eq_ct_nonlin1}-\eqref{eq_dt_nonlin2} the solution reaches the exact optimizer of \eqref{eq_dra}. In contrast, for uniform quantization, one can define $\ve$-accuracy as a trade-off between the quantization level and convergence to the $\ve$-neighborhood of the exact optimizer $\mb{x}^*$. We consider nonlinear CT protocol \eqref{eq_ct_nonlin2} and DT protocol \eqref{eq_dt_nonlin2} with $h(\partial_{x_i} f_i(k))=q\left[\dfrac{\partial_{x_i} f_i(k)}{q}\right]$ with $[\cdot]$ as rounding to the nearest integer and $q$ as the quantization level. Note that, from the definition, for $x_i$ satisfying $- 0.5 q  \prec \partial_{x_i} f_i - \partial_{x_i} f_i^* \prec 0.5 q $ we have $ h(\partial_{x_i} f_i) = h(\partial_{x_i} f_i^*)$ and for the optimizer we have $ \partial_{x_j} f_j^* = \partial_{x_i} f_i^*$. Define a new variable  $\xi(\mb{x}) := \nabla F (\mb{x}) -  \frac{\sum_{i=1}^n \partial_{x_i} f_i}{n} \mb{1}_n$. Then, from the definition, 
\begin{align}
\nabla F - \nabla F^* &=\xi +  \frac{\sum_{i=1}^n \partial_{x_i} f_i}{n}\mb{1}_n -\nabla F^* \\ \label{eq_xi1}
&= \xi +  \frac{\sum_{i=1}^n \partial_{x_i} f_i}{n}\mb{1}_n -  \frac{\sum_{i=1}^n \partial_{x_i} f_i^*}{n}\mb{1}_n
\end{align} 
where we simplified the notation as $\nabla F(\mb{x}^*) = \nabla F^*$ and $\partial_{x_i} f_i(x_i^*) = \partial_{x_i} f_i^* $.
Recall the following lemma.
\begin{lem} \label{lem_zbar}
	For ${\mb{z} \in \mathbb{R}^n}$, ${\overline{\mb{z}} := \mb{z} - \frac{\mb{1}_n^\top \mb{z}}{n} \mb{1}_n}$, and laplacian matrix  $L$ of a weight-balanced graph: $\mb{z}^\top L \mb{z} = \overline{\mb{z}}^\top L \overline{\mb{z}}$.
\end{lem} 

Putting $L=I_n$ and $\mb{z} = \nabla F - \nabla F^* $ in the above lemma along with \eqref{eq_xi1},
\begin{align}
\xi^\top \xi &= (\nabla F - \nabla F^*)^\top (\nabla F - \nabla F^*).
\end{align}
For $|\partial_{x_i} f_i - \partial_{x_i} f_i^*| < 0.5 q$ (or $|\partial_{x_i} f_i - \partial_{x_j} f_j| < q$) we have,
\begin{align} \label{eq_xixi}
\xi^\top \xi &<  0.25q^2 \mb{1}_n^\top \mb{1}_n = 0.25nq^2.
\end{align}

From Lemma~\ref{lem_strict}, substituting $ \mb{x}_1 = \mb{x}$ and $ \mb{x}_2 = \mb{x}^*$ we get,  
\begin{align} 
 \delta \mb{x}^\top \nabla F^* +  v\delta \mb{x}^\top \delta \mb{x} \leq \overline{F} \leq  \delta \mb{x}^\top \nabla F^*  +  u\delta \mb{x}^\top \delta \mb{x}
\end{align}

It is clear that for any two feasible states $\delta \mb{x}^\top \mb{1}_n = 0$ and, 
\begin{align} \label{eq_xi}
 \delta \mb{x}^\top \nabla F^* = \delta \mb{x}^\top \xi(\mb{x}^*) = 0, 
\end{align} 
since $\xi(\mb{x}^*)=\mb{0}_n$ from the definition. Further, following the results in \cite{cherukuri2015tcns} one can show that for any feasible state the residual $\overline{F}(\mb{x}) = F(\mb{x}) - F(\mb{x}^*)$ satisfies,
\begin{align} 
\label{eq_rho2}
\frac{1}{4u}     \xi^\top    \xi &\leq  \overline{F} \leq   \frac{1}{4v}     \xi^\top       \xi 
\end{align}
where we dropped the dependence on $\mb{x}$ for notation simplicity. Eq.~\eqref{eq_xi}-\eqref{eq_rho2} along with Lemma~\ref{lem_strict} result in the following.

\begin{lem} \label{lem_mine} 
	Let Assumptions~(1)-(2) hold and $2v \leq \partial_x^2 f_i(\mb{x}_i) \leq 2u$. 
	Then,
	\begin{align} 
	\label{eq_rho4} 
	v \|\mb{x} - \mb{x}^*\|^2_2 &\leq \overline{F} \leq u \|\mb{x} - \mb{x}^*\|^2_2, \\ \label{eq_rho5}
	\frac{\|\xi\|_2}{2u} &\leq \|\mb{x}-\mb{x}^* \|_2 \leq \frac{\|\xi\|_2}{2v}.
	\end{align}
\end{lem}

From \eqref{eq_rho5} and \eqref{eq_xixi} and given quantization level $q$,
\begin{align} \label{eq_ve_q}
\|\mb{x} - \mb{x}^*\|_2 &\leq \frac{\| \xi\|_2}{2v} < \frac{\sqrt{n}q}{4v}=\varepsilon. 
\end{align}
This gives an estimate that how close we can get to the optimizer $\mb{x}^*$ for uniform quantization with level $q$, i.e., the so-called $\varepsilon$-accuracy. For a given demanded accuracy level $\ve$, any quantization level  $q>\frac{4v\varepsilon}{\sqrt{n}}$ may not guarantee such $\ve$-accuracy and should be redesigned.  
One can find similar $\ve$-bound for the node-based CT protocol~\eqref{eq_ct_nonlin1} and DT protocol~\eqref{eq_dt_nonlin1} following the same line of reasoning.
\begin{rem}
	Note that the proposed nonlinear solutions are not limited to the quadratic cost model discussed in \cite{rikos2021optimal}. In general, any cost function satisfying Assumption~(1) is valid in this work. Therefore, although the consensus-based solution in \cite{rikos2021optimal} reaches the exact optimizer for quadratic costs, it is not applicable for general \textit{non-quadratic costs}. Further, the proposed solutions 
	can address penalty and barrier functions discussed in Section~\ref{sec_penalty} which are non-quadratic in general. 
	On the other hand, the proposed protocols \eqref{eq_ct_nonlin1}-\eqref{eq_dt_nonlin2} can address other types of sector-based nonlinearities with exact optimality. Solutions based on fixed-time convergent algorithms can also  be discussed as in \cite{mrd2020fast,garg_cdc20}. 
\end{rem}

\section{Possible Applications and Simulations} \label{sec_sim}
\subsection{CPU Scheduling in Data Centers}
\begin{figure} 
	\centering
	\includegraphics[width=1.65in]{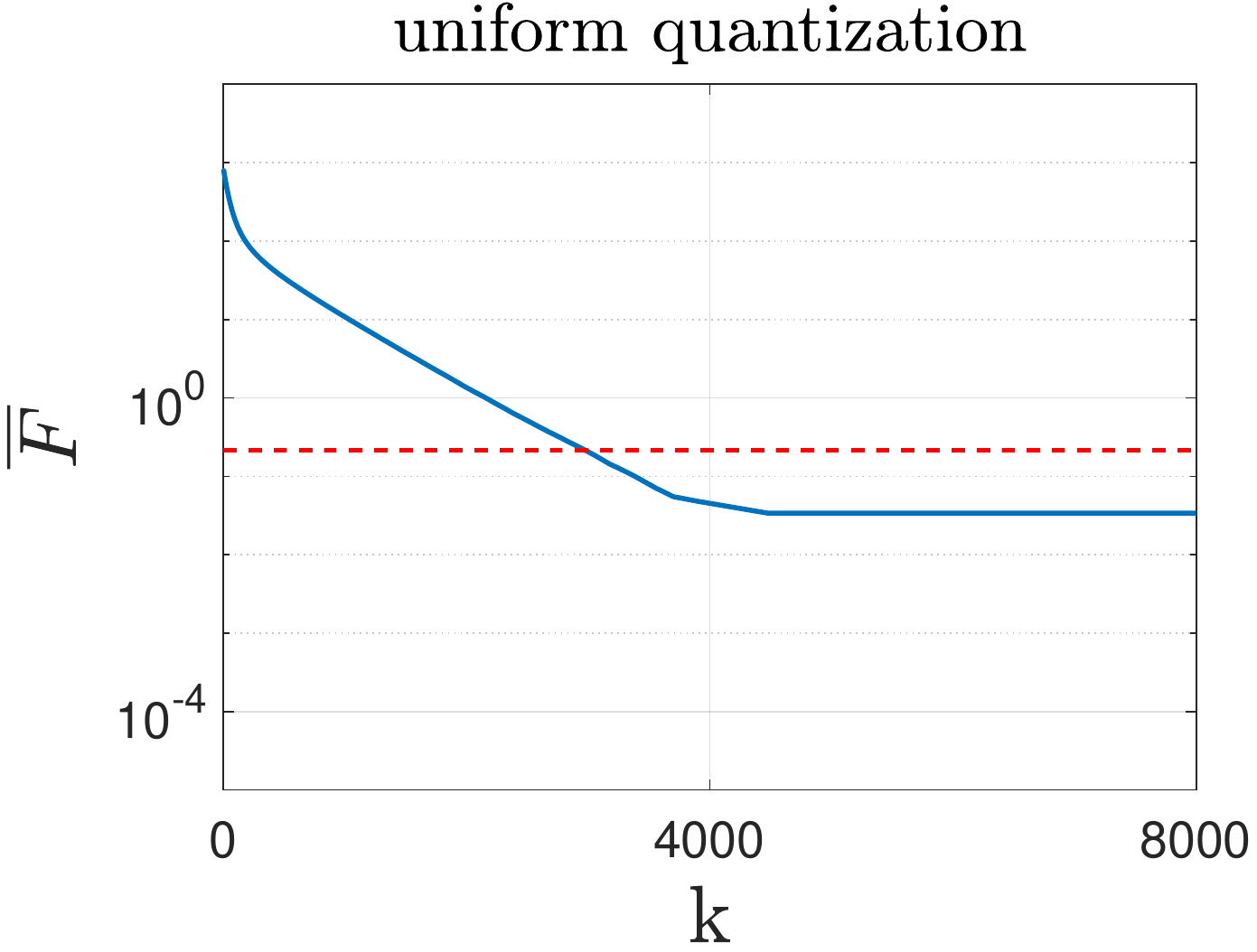}
	\includegraphics[width=1.65in]{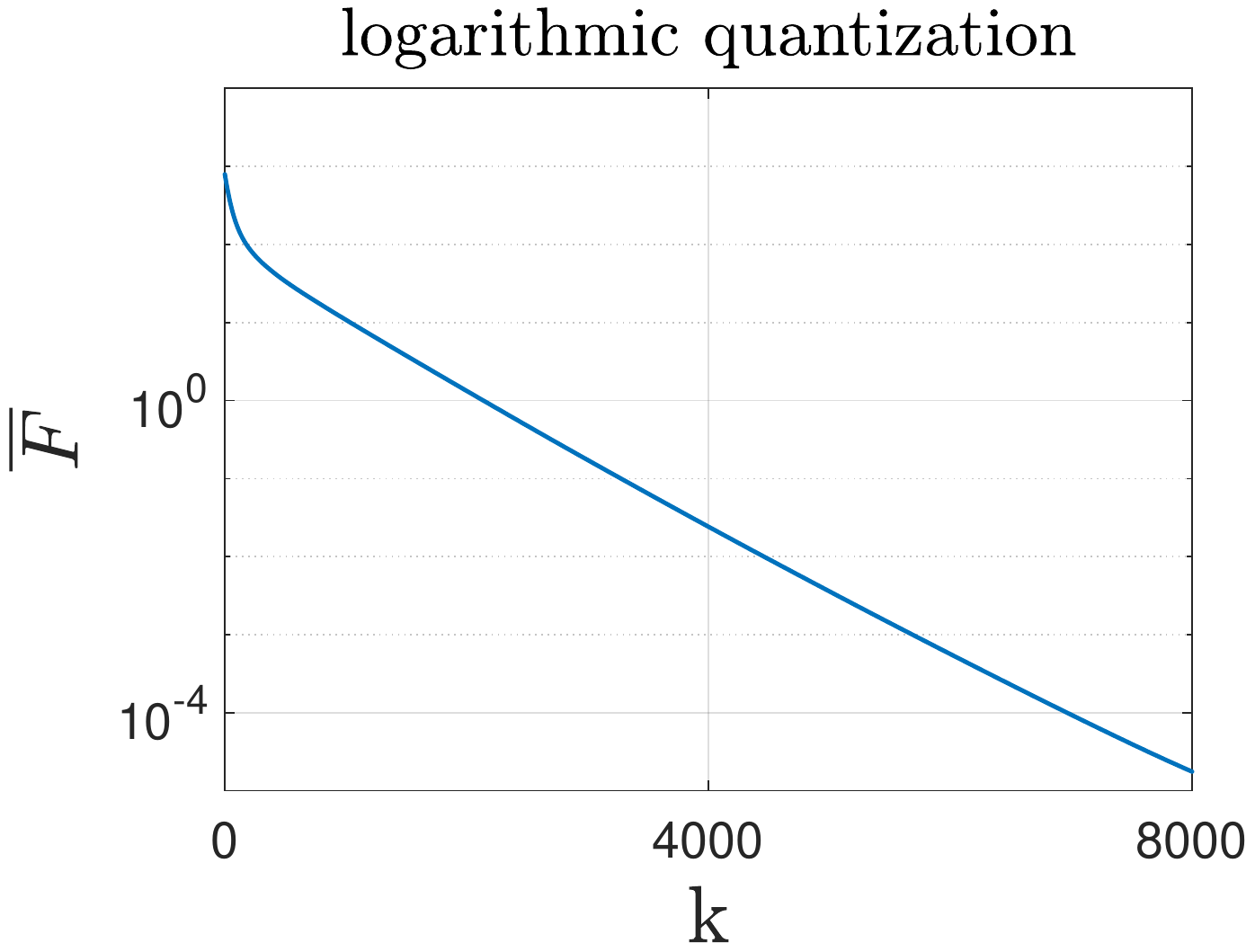}	
	\caption{The residual under two quantization approaches: (left) uniform, and (right) logarithmic quantization with level $q = 1$. Logarithmic quantizer as a sector-based nonlinearity is ''strongly'' sign-preserving as $\lim_{z \rightarrow 0} \frac{h(z)}{z}\geq(1-\frac{q}{2})>0$ and the residual converges to zero. In contrast, the uniform quantizer with $\frac{h(z)}{z}=0$ for $-\frac{h}{2} < z < \frac{h}{2}$ results in steady-state residual and converges to the $\ve$-neighborhood of the exact optimizer defined by Eq.~\eqref{eq_ve_q} and represented by the red dashed line on the left figure.  } \label{fig_quant} \vspace{-0.65cm}
\end{figure}

Consider the problem of balancing the CPU utilization over a cloud of $n=12$ data servers in order to optimally assign the CPU resources to the workloads \cite{makridis2020robust,rikos2021optimal}. The CPU costs at each node follow the quadratic form,
\begin{align}
f_i(\mb{x}_i) = \frac{1}{2} \pi_i (\mb{x}_i - \frac{\rho_i + u_i}{\pi_i})^2
\end{align}
with scalar $\pi_i > 0$ representing the capacity of node $i$, $\rho_i \in \mathbb{R}$ as the number of CPU cycles needed, and $u_i \in \mathbb{R}$ as the number
of occupied cycles due to predicted or known utilization from already running tasks on the server $i$ (see more details in \cite{makridis2020robust,rikos2021optimal}). For the simulation we choose $\pi_i =2$, random $\rho_i,u_i \in [0~ 50]$ and assume scalar box constraints on the workloads/jobs at each node as,
\begin{align}
m_i = 0 \leq \mb{x}_i \leq 100 = M_i
\end{align}
These constraints are addressed via quadratic penalty function \eqref{eq_equiv2} with $\kappa = 2$. 
Each node locally computes the optimal proportion of its workload out of $b= \sum_{i=1}^n (\rho_i+ u_i) = 563$. The communication network is considered as a simple undirected cyclic network. Let assume admissible quantization level $q = 0.125$. Substituting $v = 1$ in Eq.~\eqref{eq_ve_q}, the solution under the nonlinear (uniformly-quantized) protocol \eqref{eq_dt_nonlin1} (for sufficiently small $\overline{\eta}$) is guaranteed to reach the $\ve$-neighborhood of the optimizer $\mb{x}^*$ satisfying,
\begin{align} 
\|\mb{x} - \mb{x}^*\|_2 & < \frac{0.125\sqrt{12}}{4}=\varepsilon. 
\end{align}
Comparison between logarithmic quantization and uniform quantization is shown in Fig.~\ref{fig_quant}.

\subsection{Non-Quadratic Cost Model}
\begin{figure}
	\centering
	\includegraphics[width=1.65in]{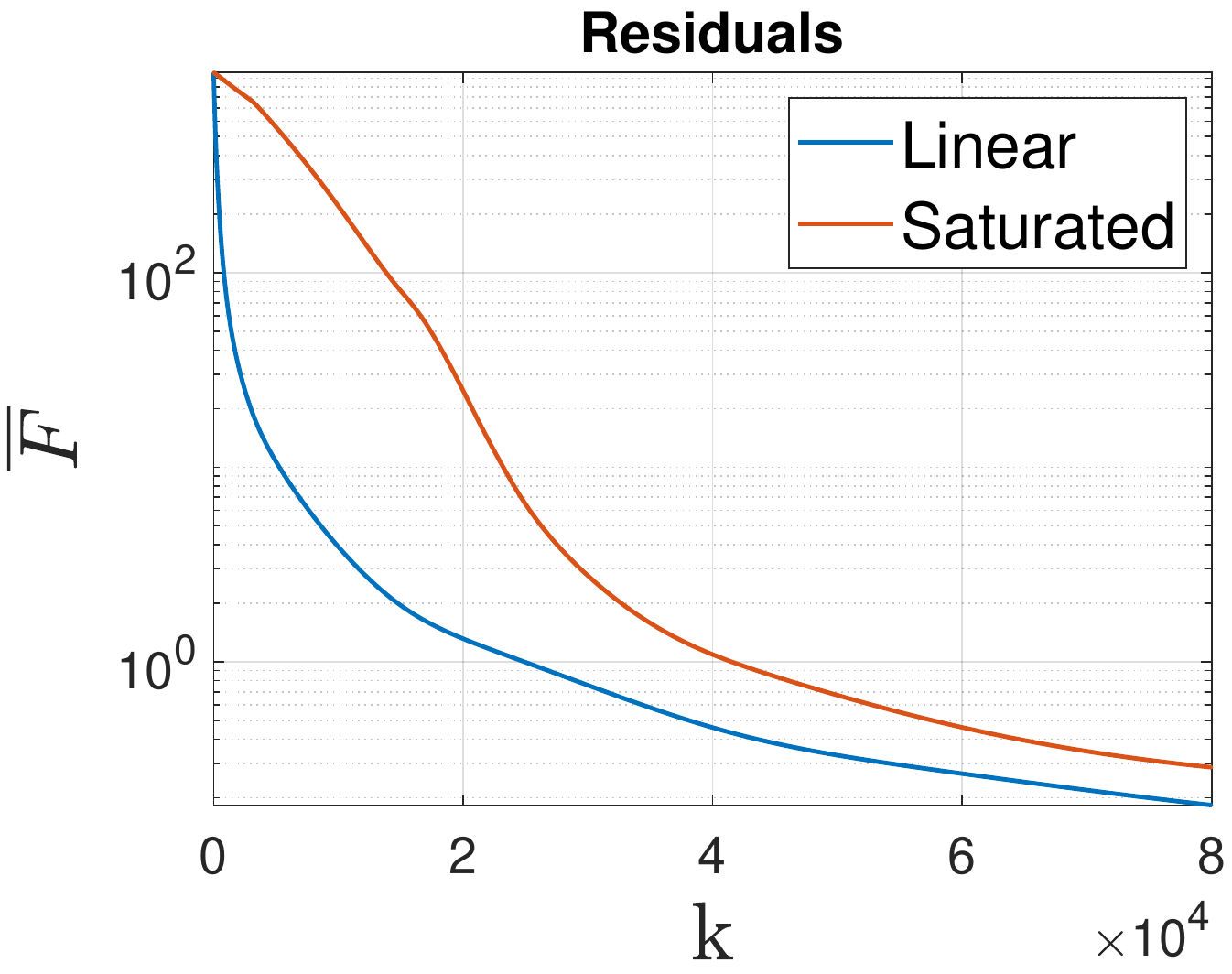}
	\includegraphics[width=1.65in]{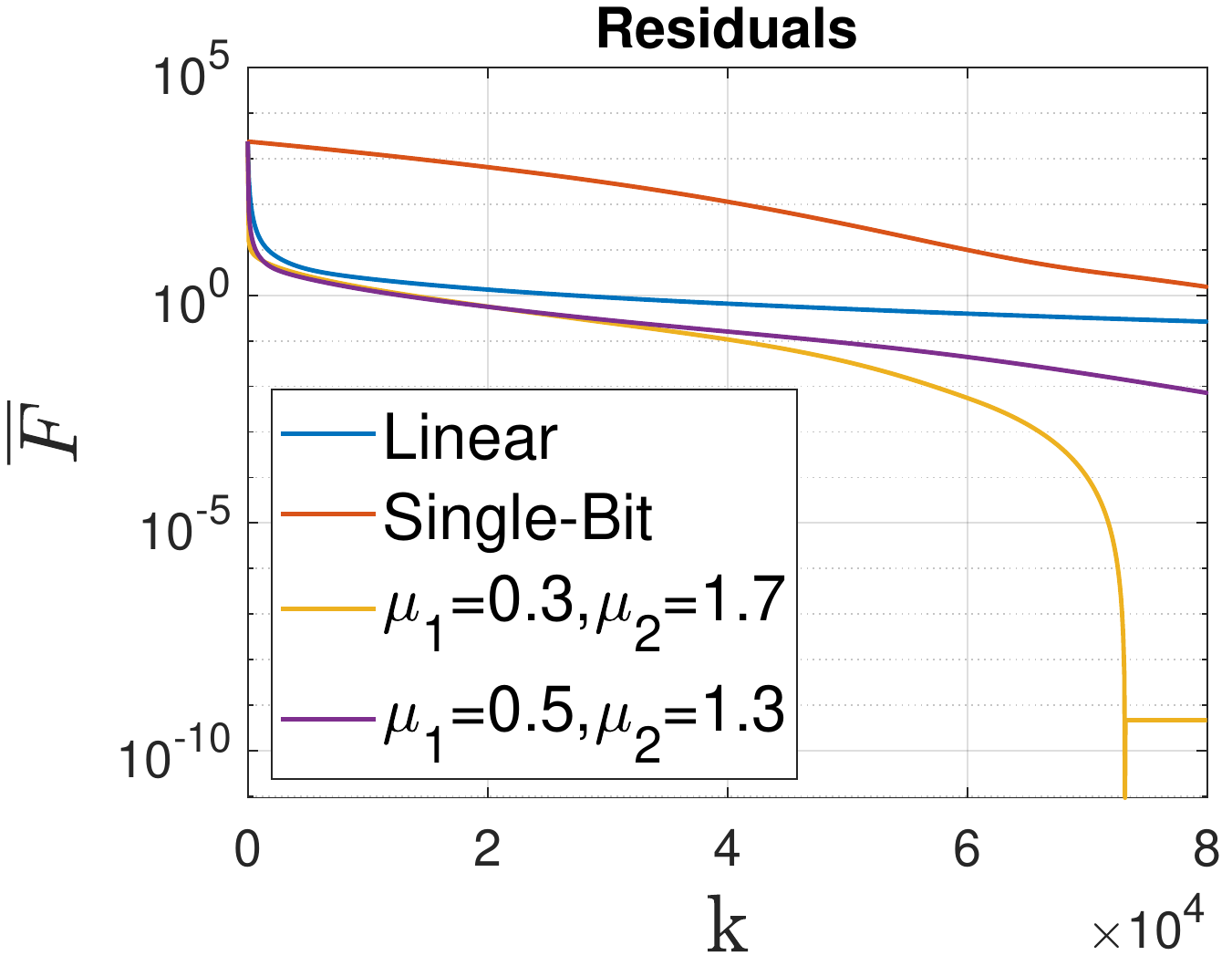}
	\caption{ (left) This figure compares the evolution of the residuals under the linear solution and the node-based protocol \eqref{eq_dt_nonlin1} subject to saturation level equal to $20$. (right) The solution under linear and different nonlinear sign-based solutions are shown. Adding sign-based nonlinearities can improve the convergence rate as compared to the linear and single-bit solutions.  } \vspace{-0.6cm}  \label{fig_satu}
\end{figure}

As mentioned in the introduction, in contrast to  consensus-based solutions that only consider quadratic cost functions \cite{rikos2021optimal}, the proposed nonlinear solution in this paper can solve resource scheduling with non-quadratic cost models.  As an example, the cost function can be in the form \cite{doan2017scl}, 
\begin{align}
\sum_{i=1}^n f_i(\mb{x}_i) = \sum_{i=1}^n \omega_i(\mb{x}_i - \alpha_i)^4
\end{align}
with random $\alpha_i \in [-2~4]$, $\omega_i \in [0~1]$. Further, the box constraints $-2 \leq \mb{x}_i \leq 5 $ can be addressed by non-quadratic (logarithmic) penalty functions \eqref{eq_equiv1} with $\mu = 1$. For this simulation, actuation saturation (protocol \eqref{eq_dt_nonlin1}) is compared with the linear solution in Fig.~\ref{fig_satu}(left). Such clipping may occur due to the maximum capacity at nodes, for example, because of some resource utilization due to previous tasks still being processed. 
In general, linear dynamics to solve the resource allocation converge slowly and asymptotically. To improve the convergence rate and to reach fixed-time convergence, sign-based solutions can be adopted. It is known that nonlinear dynamics in the form $\dot{\mb{z}} = \mbox{sgn}^{\mu_1}(\mb{z}) + \mbox{sgn}^{\mu_2}(\mb{z})$ converge to the equilibrium in fixed (or prescribed) time \cite{garg_cdc20}. Choosing nonlinear function $h(\mb{z})=\mbox{sgn}^{\mu_1}(\mb{z}) + \mbox{sgn}^{\mu_2}(\mb{z})$ one can improve the convergence rate of the proposed protocols \eqref{eq_ct_nonlin1}-\eqref{eq_dt_nonlin2} to reach faster convergence as compared to the existing linear solutions \cite{boyd2006optimal}.
The simulation results are shown in Fig.~\ref{fig_satu} for two cases with $\mu_1 = 0.5,\mu_2 = 1.3$ and $\mu_1 = 0.3,\mu_2 = 1.7$ for protocol \eqref{eq_dt_nonlin1}  along with the single-bit protocol by \cite{taes2020finite} (with $\eta = 3 \times 10^{-5} $). Due to non-Lipschitz continuity of the sign-based solutions, in \textit{discrete-time}, the steady-state residual is biased (known as the so-called \textit{chattering} phenomena). This bias can be reduced by decreasing the step rate $\eta$. 

\section{Discussions and Concluding Remarks} \label{sec_con}
This paper considers node-based and link-based nonlinearities on the agents' dynamics to optimally solve resource allocation subject to global sum-preserving constraints and local box constraints. In particular, the application to CPU scheduling subject to logarithmic quantization (sector-based nonlinearity) and uniform quantization  (non-sector-based nonlinearity) are compared and for the latter $\ve$-accuracy is addressed.  
As an extension and future research direction, the higher-order state dimension at agents  can be considered as,
 \begin{eqnarray} \label{eq_dra_A}
 \min_\mb{y}
 ~ & \widetilde{F}(\mb{y}) = \sum_{i=1}^{n} \widetilde{f}_i(\mb{y}_i)\\ \nonumber
 & \text{s.t.} ~~ \sum_{i=1}^{n} A_i\mb{y}_i = \mb{b} \\  \nonumber
 & \mb{y}_i \in \mc{Y}_i
 \end{eqnarray}
 with $\mb{y}_i \in \mathbb{R}^{d_i}, \mb{b} \in \mathbb{R}^{m} $, $\widetilde{f}_i: \mathbb{R}^{d_i} \rightarrow \mathbb{R}$, $\mc{Y}_i \subseteq \mathbb{R}^{d_i}$, and $A_i \in \mathbb{R}^{m \times d_i}$ as a full row-rank matrix. Note that the feasibility constraint $ \sum_{i=1}^{n} A_i\mb{y}_i = \mb{b}$ is the summation of some local constraints (of higher dimension).
 One point to notice is the convexity of the local constraints to admit certain composition conditions as discussed in \cite[Section~3.2.4]{boyd2004convex}. Such extensions based on the results of \cite{mikael2021cdc} can be addressed as a promising direction of future research.   

\bibliographystyle{IEEEbib}
\bibliography{ifacbib}
\end{document}